\RequirePackage[2020-02-02]{latexrelease}
\documentclass[%
 preprint,
 superscriptaddress,
 amsmath,
 amssymb,
 aps,
 prl
]{revtex4-1}

\usepackage{graphicx}
\usepackage{dcolumn}
\usepackage{bm}
\usepackage[colorlinks,citecolor=blue]{hyperref}

\usepackage{xspace}
\usepackage{siunitx}

\newcommand{\fig}[1]{Fig.~\ref{#1}}
\newcommand{\figu}[1]{Figure~\ref{#1}}

\newcommand{\hrms}{\ensuremath{h_\mathrm{rms}}\xspace}

\newcommand{\etal}{et al.\xspace}
\newcommand{\cuzr}{Cu\textsubscript{50}Zr\textsubscript{50}\xspace}

\newcommand{\pixeltag}{\ensuremath{\mathrm{pix}}}

\newcommand{\lenpix}{\ensuremath{L_{\pixeltag}}\xspace}

\newcommand{\psdtwodprefix}{C^\mathrm{2D}}
\newcommand{\psdisoprefix}{C^\mathrm{iso}}
\newcommand{\pixpsdtwod}{\ensuremath{\psdtwodprefix_{q_xq_y}}\xspace}
\newcommand{\pixpsdiso}{\ensuremath{\psdisoprefix_{q}}\xspace}

\newcommand{\pixhft}{\ensuremath{\tilde{h}_{q_xq_y}}\xspace}
\newcommand{\pixh}{\ensuremath{h_{xy}}\xspace}

\begin{document}

\title{Nonequilibrium plastic roughening of metallic glasses yields self-affine topographies with strain-rate and temperature-dependent scaling exponents}

\author{Wolfram G. N\"ohring}
\affiliation{Department of Microsystems Engineering, University of Freiburg, 79110 Freiburg, Germany}

\author{Adam R. Hinkle}
\affiliation{Institute for Applied Materials, Karlsruhe Institute of Technology, 76131 Karlsruhe, Germany}
\affiliation{DCS Corporation, Belcamp, MD 21017, USA}
\affiliation{U.S. Army Combat Capabilities Development Command, Chemical and Biological Center, Aberdeen Proving Ground, MD 21010-5424, USA}

\author{Lars Pastewka}
\affiliation{Department of Microsystems Engineering, University of Freiburg, 79110 Freiburg, Germany}
\affiliation{Institute for Applied Materials, Karlsruhe Institute of Technology, 76131 Karlsruhe, Germany}
\affiliation{Cluster of Excellence livMatS, Freiburg Center for Interactive Materials and Bioinspired Technologies, University of Freiburg, 79110 Freiburg, Germany}

\date{\today}

\begin{abstract}
We study nonequilibrium roughening during compressive plastic flow of initially flat Cu$_{50}$Zr$_{50}$ metallic glass using large-scale molecular dynamics simulations. Roughness emerges at atomically flat interfaces beyond the yield point of the glass. A self-affine rough topography is imprinted at yield and is reinforced during subsequent deformation. The imprinted topographies have Hurst exponents that decrease with increasing strain-rate and temperature. After yield, the root-mean-square roughness amplitude grows as the square-root of the applied strain with a prefactor that also drops with increasing strain-rate and temperature. Our calculations reveal the emergence of spatial power-law correlations from homogeneous samples during plastic flow with exponents that depend on the rate of deformation and the temperature. The results have implications for interpreting and engineering roughness profiles.
\end{abstract}


\maketitle

Roughness controls many interfacial phenomena. The
most prominent examples are arguably the influence of roughness on friction~\cite{urbakh_nonlinear_2004,persson_roughness} and
adhesion~\cite{fuller_effect_1975,maugis_contact_1996,persson_effect_2001,persson_adhesion_2002,pastewka_contact_2014,dalvi_linking_2019}. It is therefore useful to characterize
surface roughness, and to understand the mechanisms that produce it, including growth, erosion, and fracture. Here, we focus on plastic
deformation as a roughening mechanism: When a solid is deformed irreversibly
by external forces, a signature of the deformation process in imprinted on the surface. For example, crystal dislocations in metals~\cite{miura_punchedout_1972,zaiser_self-affine_2004,gagel_formation_2016,gagel_discrete_2018,hinkle_emergence_2020} or shear bands and fractures in rocks~\cite{fossen_structural_2016} leave slip traces behind that roughen surfaces.

Unlike crystalline metals, the fundamental deformation event in glasses is not a slip trace but a localized shear transformation or STZ (shear transformation zone) ~\cite{spaepen_microscopic_1977,argon_plastic_1979,falk_dynamics_1998,hufnagel_deformation_2016}. At low temperature (or slow deformation), these can coalesce to form deformation avalanches that potentially span the size of the whole system~\cite{lemaitre_rate-dependent_2009,Hentschel_size_2010}. Despite the importance of glasses for engineering applications and the wide interest in deformation of glasses in physics, little is known about the nonequilibrium processes forming their surface morphology.

Glasses formed by quenching have surfaces that are well described by capillary waves~\cite{jackle_intrinsic_1995,sarlat_frozen_2006,zhang_roughness_2021}. However, many natural and industrially processed surfaces are found to be self-affine fractals with power-law scaling of heights
\cite{sayles_surface_1978,mandelbrot_fractal_1982,mecholsky_quantitative_1989,bonamy_scaling_2006,candela_roughness_2012,gujrati_combining_2018}.
The root mean square (rms) amplitude of surface height fluctuations, $\hrms(L)$, measured in a square region with side of length $L$, then scales as $\hrms(L) \propto L^{H}$ where $H$ is the Hurst exponent.

Recent simulation studies have shown that deformation
is one possible origin of self-affinity. For example, Milanese \etal{}
\cite{milanese_emergence_2019}
have observed self-affine roughening in 2D discrete-element simulations
of sliding contacts that formed abrasive third bodies. We have recently
shown using molecular-dynamics calculations that self-affine roughness emerges naturally
during compression of atomically
flat surfaces made from pure metals, crystalline alloys,
and metallic glasses~\cite{hinkle_emergence_2020}. Similar results have been obtained by Vacher \& de Wijn for the surface of polymers~\cite{vacher_molecular-dynamics_2021}.

We here present evidence that -- unlike in crystalline materials~\cite{hinkle_emergence_2020} -- the roughness characteristics of a glass depend
strongly on temperature and deformation rate. We
show that two regimes of temperature and strain rate must be distinguished.
In quasi-static deformation, and at the lowest temperatures and rates,
the topography is dominated by system-spanning shear bands. Self-affine scaling of surface heights is nevertheless plausible through a mechanism similar to slip of crystal dislocations, but with a shear band as the fundamental slip event.
Higher temperatures and strain rates lead to the formation of smoother and more
diffuse topographies that exhibit self-affine scaling at small scales. Their rms roughness \hrms grows roughly as
$\hrms\propto\varepsilon^{1/2}$ at large strain.
Our simulations show that the topography is imprinted shortly after yield and is reinforced during subsequent deformation.

\begin{figure}[htb]
    \includegraphics{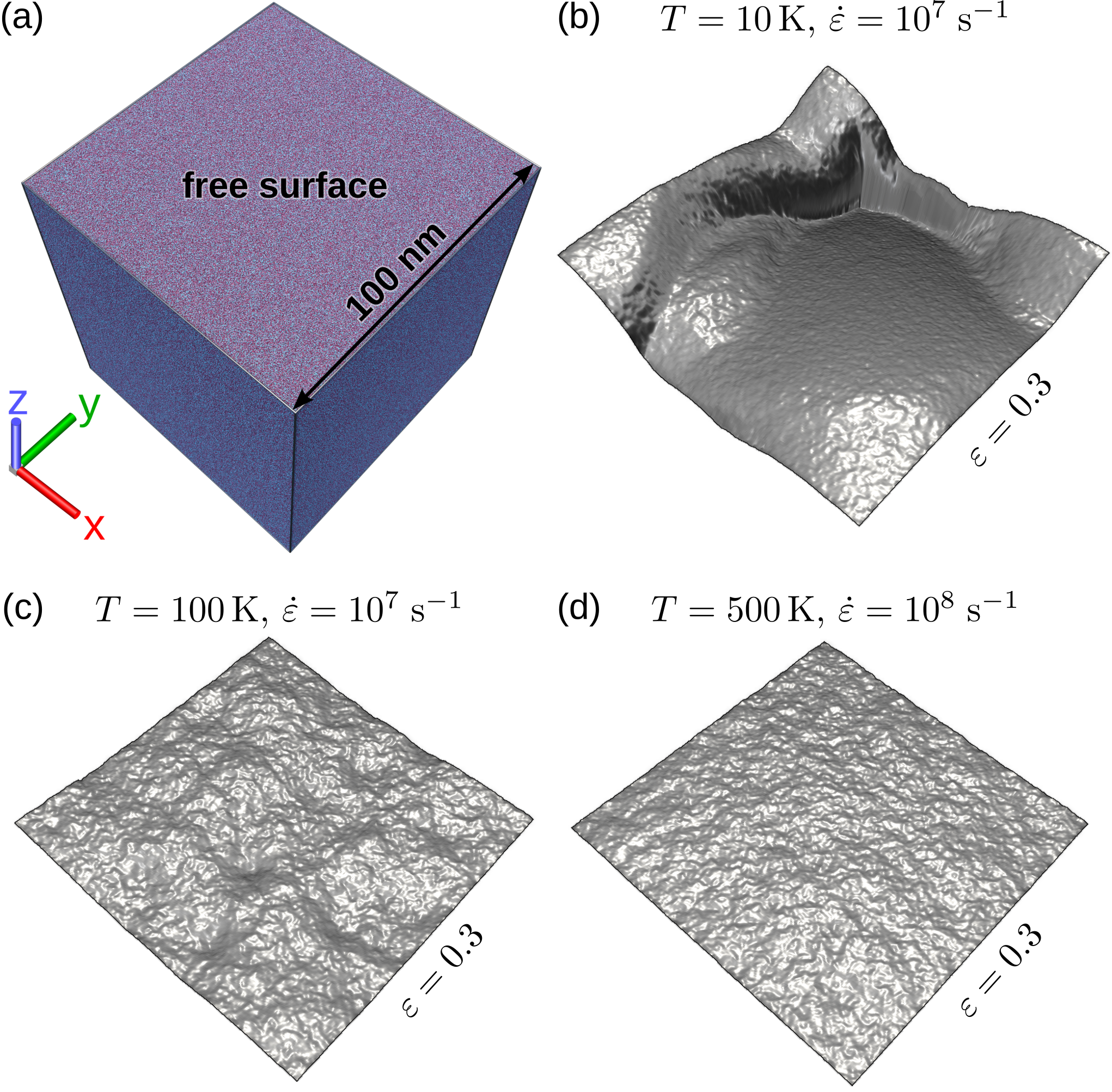}
    \caption{\label{fig:one}(a)
        \cuzr cube with a side length of \SI{100}{\nano\meter} and a free surface in the $z$-direction. We simulated biaxial
        compression along the $x$- and $y$-directions. (b) Topography
        of the surface after \SI{30}{\percent}
        compression with a rate of \SI{e8}{\per\second} at \SI{10}{\kelvin}.
        A shear band has formed. No such distinct feature is visible at
        (c) \SI{100}{\kelvin} and (d) \SI{500}{\kelvin}. The topography
        becomes smoother with increasing temperature.
    }
\end{figure}

To study the emergence of surface roughness, we simulated \SI[product-units = single]{100 x 100 x
100}{\nano\metre} cubes of amorphous \cuzr using molecular dynamics. The initial configuration shown in \fig{fig:one}a, contained \SI{5.8e7}{} atoms, interacting with the embedded atom method potential
of Cheng \& Ma \cite{cheng_atomic-level_2011}.
The initial structure was melt-quenched
at a rate of \SI{e11}{\kelvin\per\second}, followed by equilibration for \SI{100}{\pico\second} at
the target temperature and zero pressure. The systems were fully periodic during quench and equilibration, after which we opened the boundaries along the
$z$-direction, creating two free surfaces. The samples were equilibrated for \SI{100}{\pico\second} after creating the free surfaces and before compression.
 
We compressed these samples by dynamically reducing the lengths of the
simulation cell in the $x$- and $y$-directions, keeping the cross-section in the $xy$-plane square. We report results in terms of the applied engineering strain $\varepsilon(t)=1-L(t)/L_0$, where
$L(t)$ is the linear length of the cell (at time $t$), and $L_0$ the initial initial length.
Since a jump in the strain rate generates undesirable shock waves,
we gradually increased $\dot{\varepsilon}(t)$ from zero to a constant
value during the first $\SI{100}{\pico\second}$ of
the simulation.
We carried out simulations at final strain rates of
$\text{\SIlist{e7;e8}{\per\second}}$.

During deformation, we controlled the
temperature using a dissipative particle dynamics (DPD)~\cite{soddemann_dissipative_2003} thermostat with a cutoff radius of \SI{6.5}{\angstrom} -- equal to the cutoff radius of the potential -- and a damping parameter
$\gamma=\SI{5e-3}{\electronvolt\pico\second\per\angstrom\squared}$, which
leads to a characteristic damping time of approximately \SI{1}{\pico\second}.
Unlike Langevin or Nos\'e-Hoover thermostats, the Gallilean-invariant DPD thermostat does not overdamp long-wavelength modes, which is important for large-scale simulations such as those reported here.

Plastic deformation roughened the initially flat sample surfaces, which we quantified from a pixel representation $h_{xy}(\varepsilon)$ of
the topography with lateral pixel size $\lenpix\approx\SI{3}{\angstrom}$. $h_{xy}$ is the height of the topmost atom (bottommost for bottom surface) associated
with the pixel with center coordinates $x$ and $y$. 
We performed all statistical analysis on $h_{xy}$, e.g., the root-mean-square roughness $\hrms = \langle h_{xy}^2 \rangle$ where $\langle\cdot\rangle$ is the average over all pixels. We assume that all heights are leveled such that $\langle h_{xy} \rangle=0$.

\fig{fig:one}b-d show \pixh after \SI{30}{\percent} compression
in our dynamic tests. At the lowest rate and temperature (\SI{e7}{\per\second}
and \SI{10}{\kelvin}, \fig{fig:one}b), the most prominent feature is a
lentil-shaped dip formed by shear bands. We find a similar topography in a quasistatic simulation, where we imposed
compressive strain on the cell in increments of
$\Delta\varepsilon=\SI{e-4}{}$ by affinely remapping all coordinates, followed by subsequent relaxation down to a force of \SI{e-3}{\electronvolt\per\angstrom} using the
\textsc{fire} algorithm \cite{bitzek_structural_2006}. In the other simulations at higher temperature or higher rate,
deformation was less localized and the resulting surface topography is more
diffuse, see \fig{fig:one}c and d. The roughness amplitude generally decreases with
increasing temperature and rate. At
$T=\SI{500}{\kelvin}$ and $\dot\varepsilon=\SI{e8}{\per\second}$, the
height range  is only \SI{2}{\nano\meter}.

\begin{figure*}
    \includegraphics{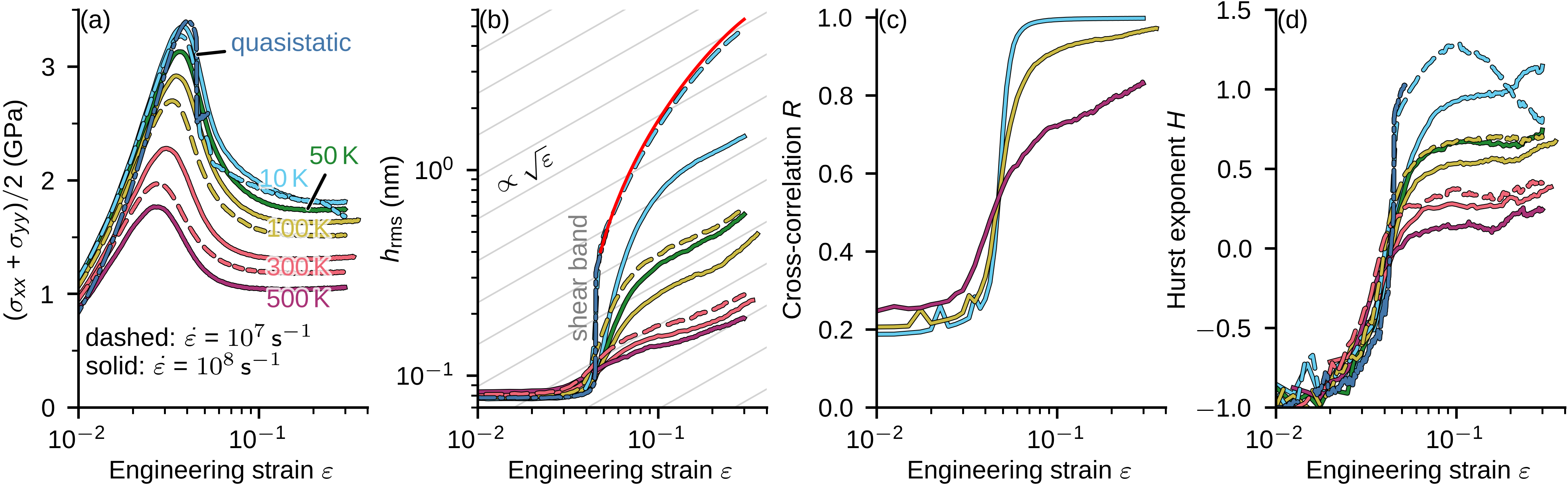}
    \caption{\label{fig:two}(a) Lateral normal stress
        vs. compressive strain $\varepsilon$ in
        simulations with different strain rates $\dot\varepsilon$
        and temperatures $T$. (b) Root-mean square roughness \hrms, which increases
        after yield. A shear band forms in the quasistatic
        simulation, and in low-rate, low-temperature simulations (here 
        $\dot{\varepsilon}=\SI{e7}{\per\second}$ at $\SI{10}{\kelvin}$).
        The thin red line is a model of \hrms for a
        topography dominated by a shear band (see text).
        (c) Cross-correlation
        $R$ (see text) between
        subsequent topographies at intervals of $\Delta
        \varepsilon\approx{}0.0025$. (d) Hurst exponent extracted from fits to the PSD (see text).
    }
\end{figure*}

\figu{fig:two}a
shows the mean lateral stress $\left(\sigma_{xx}+\sigma_{yy}\right)/2$ as a function of strain $\varepsilon$.
In most cases, the elastic regime does not end abruptly. After the peak, the
stress decreases smoothly, as expected~\cite{rottler_shear_2003,bhowmick_plastic_2006,shimizu_yield_2006},
and stabilizes at the steady-state flow stress. Increasing temperature and
decreasing rate lowers the peak and flow stresses. The only exceptions are the two simulations with shear banding: the
quasistatic test and the dynamic simulation with
$\dot\varepsilon=\SI{e7}{\per\second}$ at \SI{10}{\kelvin}. In the former
case, a sharp stress drop is visible. In the latter case, the stress decreases
again close to $\varepsilon=0.3$. 

In all cases, \hrms (see Fig.~\ref{fig:two}b) is initially less than
$\SI{0.1}{\nano\meter}$, as should be expected for an atomically flat
surface, but increases after peak stress. The rate of roughening
depends  on $\dot\varepsilon$ and $T$. The steepest
increase is seen in the quasistatic simulation and the simulation  with
$\dot\varepsilon=\SI{e7}{\per\second}$ at \SI{10}{\kelvin}.
Increasing $\dot\varepsilon$ and $T$ leads to a more gradual
transition, and lower \hrms at the same strain. At the highest temperatures, there is little
difference between the curves for \SI{300}{\kelvin} and \SI{500}{\kelvin}.
At large strain, $\hrms\propto\varepsilon^\alpha$ with
$\alpha\approx{}0.5$, except in the simulation  with
$\dot\varepsilon=\SI{e7}{\per\second}$ at \SI{10}{\kelvin} that forms a shear band.

While \hrms increases continuously, there is little qualitative difference
between the topography formed during yield, and the topography at later stages
of deformation. To quantify this observation, we calculated
the cross-correlation $R(\varepsilon)=\langle h_{xy}(\varepsilon-\Delta\varepsilon) h_{xy}(\varepsilon) / [\hrms(\varepsilon-\Delta \varepsilon)\hrms(\varepsilon)]\rangle^{1/2}$ between subsequent
simulation snapshots with $\Delta\varepsilon=0.0025$ for the simulations with
$\dot\varepsilon=\SI{e8}{\per\second}$ (\fig{fig:two}c).
Before yield, the
cross-correlation is small ($\approx{}0.2$), indicating
that there are not many common features between subsequent
snapshots. This is not surprising, since \hrms
is so low that thermal fluctuations dominate the topography. After
the
peak stress, the cross-correlation increases significantly,
with the strongest and most rapid increase seen in the simulation
at the lowest temperature of $T=\SI{10}{\kelvin}$. In this case,
the cross-correlation jumps to a value close to one, indicating that the
topography is reinforced -- peaks grow
and valleys become deeper -- with little qualitative change.
Increasing temperature reduces the correlation, but beyond
$\varepsilon\approx{}0.04$, it stays above $0.5$.

\begin{figure}
    \includegraphics{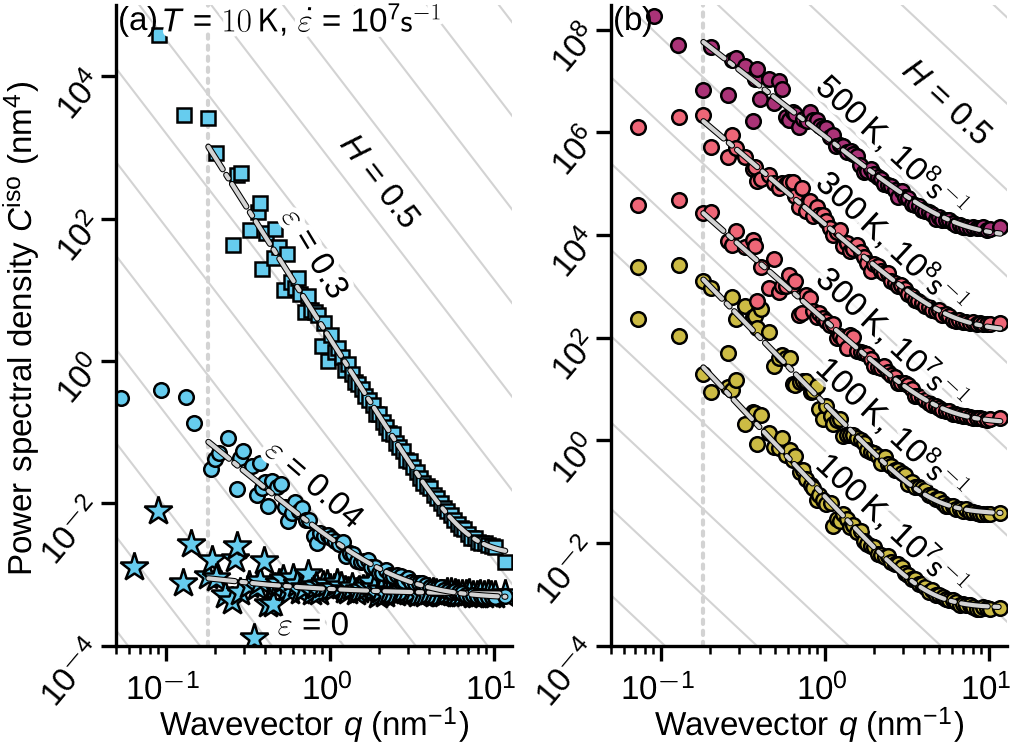}
    \caption{\label{fig:three}(a) Radially averaged power spectral
        density (PSD) $\pixpsdiso$ of the surface heights,
        at three different strains in the simulation with
        $T=\SI{10}{\kelvin}$ and $\dot\varepsilon=\SI{e7}{\per\second}$. As
        a consequence of plastic deformation, $\pixpsdiso$
        increases with strain $\varepsilon$, first at long wavelengths (small wavevectors $q$). Despite a shear band that forms in this simulation,
         $\pixpsdiso$ is approximately linear in the double
         logarithmic plot at large $\varepsilon$. (b)  $\pixpsdiso$
         at $\varepsilon=0.3$ in simulations at and above $100$~K. 
         Gray lines: Fit to a white noise plus power-law model $\pixpsdiso= C_\text{wn}+C_\text{f}q^{-2-2H}$ in 
         the region $q>\SI{0.18}{\per\nano\meter}$; the cutoff is indicated by a vertical dashed line. The ordinates in (b) have been shifted for better visibility.
    }
\end{figure}

In order to examine lateral correlation in the topography,
we compute the power-spectral density (PSD), $\pixpsdtwod  = L^{-2}\vert \pixhft \vert^2$ (or rather its radial average
 $\pixpsdiso$),
where \pixhft is the discrete Fourier transform of \pixh. (See Ref.~\cite{jacobs_quantitative_2017} for the conventions used here.) If the topography is self-affine with Hurst exponent $H$, then the
PSD scales as $\pixpsdiso \propto  q^{-2-2H}$. \fig{fig:three}a shows how \pixpsdiso evolves with
strain $\varepsilon$. The PSD is
constant at small strain, where the residual roughness is
given by uncorrelated thermal noise and quenched disorder of
the glassy state.
At yield, \pixpsdiso begins to grow, as can be seen in
the curve for $\varepsilon=0.04$, which is just past the peak stress in this simulation.
Topographic structure emerges first at small $q$ or long wavelengths.
At a large strain of  
$\varepsilon=0.3$, \pixpsdiso has a linear region in the double-logarithmic
plot, which is the signature of self-affine (scale-free)
roughness.

\fig{fig:three}b shows \pixpsdiso from the
simulations at \SIlist{100;300;500}{\kelvin}, at $\varepsilon=0.3$. In the double-logarithmic plot, 
all curves have a linear region at intermediate $q$. As a guide to the eye, the gray lines in Figs.~\ref{fig:three}a and b show ideal fractal scaling with a Hurst exponent $H=0.5$. We extract an estimate for the Hurst exponent from these calculations by fitting the simple model $\pixpsdiso = C_\text{wn} + C_\text{f} q^{-2 -2 H}$ to the PSD data, excluding data in the long wavelength region 
$q<\SI{0.18}{\per\nano\meter}$. The constant $C_\text{wn}$ is a white noise contribution and $C_\text{f}$ the amplitude of the fractal regime.

The individual fits are shown as dashed gray lines in \fig{fig:three}a and b.
The combination of white noise and a self-affine regime describe the
data well, even at at $\varepsilon=0.04$ in \fig{fig:three}a 
where roughening has just begun. The fits now allow us to plot the evolution of $H$ with strain $\varepsilon$ (\fig{fig:two}d). There is an initial region of negative $H$ where the surfaces are flat and do not show self-affine scaling, followed by a jump to a finite value as the surfaces yield. $H$ depends on temperature and strain rate, with lower temperatures and lower strain rates leading to larger values of $H$.

At low temperature and strain rate, the system is near
the athermal quasistatic (or overdamped) regime of deformation~\cite{salerno_avalanches_2012}, with the quasistatic simulation as the limiting case. In line with many previous
investigations~\cite{albe_enhancing_2013,Singh_brittle_2020}, we observe system-spanning shear bands that form topography by leaving steps on the surface.
The corresponding $\hrms$-curve in \fig{fig:two}b
can thus be understood using a simple model of the growth of a single
surface step described by the function $h(x) = a \left(x - L(\varepsilon)/2\right)$, see Supplemental Material. 
The result is shown by the red line in \fig{fig:two}b. A random distribution of such steps then leads to a self-affine topography at scales larger than our simulations (see also discussion on dislocations in Ref.~\cite{hinkle_emergence_2020}).

It is remarkable that the topography appears to be self-affine
at large strain (\fig{fig:three}a), even though the overall topography is dominated a single system-spanning shear
band. We note that an idealized sawtooth
profile (that is \emph{not} self-affine) also exhibits power-law scaling
$\pixpsdiso\propto{}q^{-3}$ of the power spectral density -- however with a smaller apparent Hurst
exponent of $H=0.5$, while our fit of
the data in \fig{fig:three}a yields $H=0.8$. Since power-law
scaling of \pixpsdiso is also plausible in the other simulations
with higher rates and temperatures, where no system-spanning
shear bands are formed, these shear bands likely do not
control the exponent of \pixpsdiso in \fig{fig:three}a.

Decreasing either rate or temperature increases
both \hrms and $H$. In this respect, the glass is different
from metal crystals, where \hrms and $H$ due to
roughening by plastic deformation are independent of
rate and temperature \cite{hinkle_emergence_2020}.
We note that in particular that the dependence of the scaling exponent $H$ on rate and temperature is unusual, as scaling exponents in power-law correlated data of phase transitions or critical phenomena are widely regarded as universal~\cite{stanley_scaling_1999}. These observations touch upon an ongoing discussion regarding whether the yielding transition (in glassy materials) can be strictly regarded as a phase transition~\cite{jaiswal_mechanical_2016,parisi_shear_2017,ozawa_random_2018,jana_correlations_2019,ozawa_role_2020}. Since our simulations suggest that $H$ is fixed at yield, the interpretation of yielding as a phase transition may only apply in the athermal case.

The differences between our topographies appear to reflect the temperature and rate sensitivity of plastic flow in the glass. A finite rate and temperature limit the magnitude of plastic events~\cite{lemaitre_rate-dependent_2009,Hentschel_size_2010}.
At fast deformation rates, new events are nucleated before the avalanches triggered
by earlier events can finish, and at high temperatures, large avalanches are overwhelmed by thermal noise because of subextensive scaling of the avalanche with system size. This is directly reflected in structural measures of flow. Different scaling relations have been reported for the scaling of a correlation length $\xi$ between plastic flow events with shear rate $\dot\varepsilon$, such as $\xi\propto\dot\varepsilon^{-1}$ (distance $\xi$ between shear bands in 3D~\cite{Singh_brittle_2020}), $\xi\propto\dot\varepsilon^{-0.4}$ (characteristic correlation length $\xi$ in 2D~\cite{clemmer_criticality_2021}) or $\xi\propto\dot\varepsilon^{-0.3}$ (characteristic correlation length $\xi$ in 3D~\cite{clemmer_criticality_2021}).

\begin{figure}
    \includegraphics{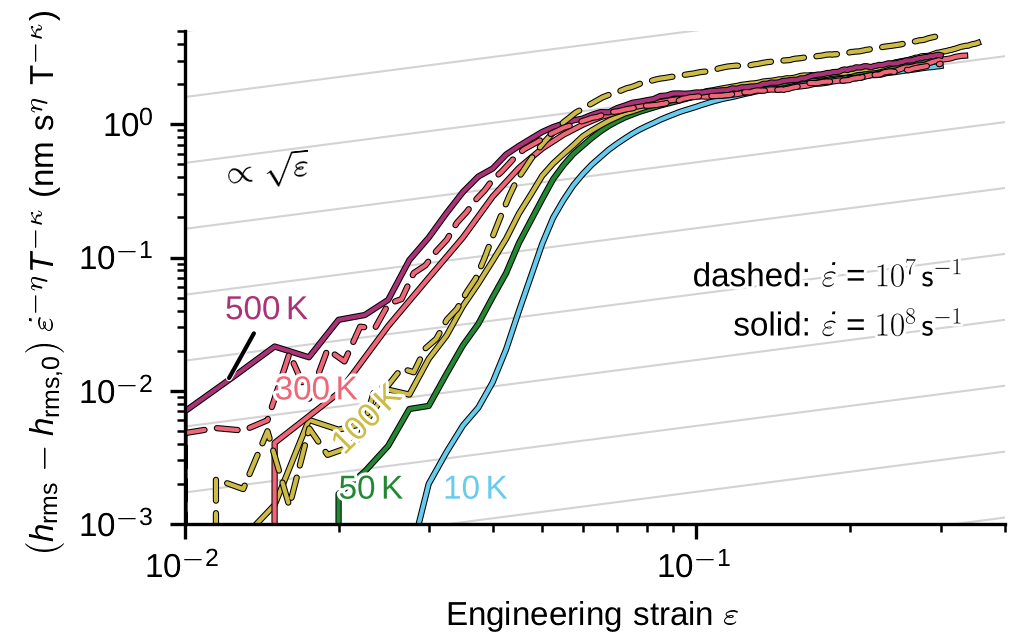}
    \caption{\label{fig:four}Root-mean square roughness \hrms,
    divided by $\gamma^{\eta}T^{\kappa}$, with $\eta=-0.1$ and
    $\kappa=-0.7$. This normalization collapses the \hrms-curves
    close at high strain.}
\end{figure}

We now attempt a similar scaling collapse of our data for the characteristic length in our system, the height amplitude \hrms. 
We attempt to collapse the $\hrms(\varepsilon)$-curves
from those simulations where no system-spanning shear band is nucleated
assuming the empirical relation $\hrms\propto\dot\gamma^{\eta}T^{\kappa}$. Before dividing
by $\dot\gamma^{\eta}T^{\kappa}$, we subtracted the small baseline
roughness $h_{\text{rms},0}$ of the undeformed state that reflects thermal fluctuations.
We computed $h_{\text{rms},0}$ as the mean value of $\hrms(\varepsilon)$ for $\varepsilon<0.01$.
\fig{fig:four} shows the curves after normalization with $\eta=-0.1$ and
$\kappa=-0.7$. This choice collapses the data beyond the yield point ($\varepsilon \gtrsim 0.1$).

In summary, the evolution of the surface roughness of a deformed Cu$_{50}$Zr$_{50}$ metallic glass reflects the dependence of plastic flow on rate $\dot\varepsilon$ and temperature $T$.
At low values of these parameters, the topography is dominated by system-spanning shear bands. At higher temperatures and rates, a more diffuse topography emerges, with some traces of universality: At large strain, the root mean square roughness
tends to grow as $\hrms\propto\varepsilon^{\alpha}$, with $\alpha\approx{}0.5$,
and lower values at the highest rates and temperatures. Moreover,
$\hrms\propto\dot\gamma^{\eta}T^{\kappa}$ at large strain with $\eta\approx-0.1$ and $\kappa\approx-0.7$. 
Regardless of rate and temperature, the power-spectral density of all
surfaces can be described as the superposition of a self-affine part and
constant noise from thermal fluctuations and quenched disorder.

Our results remain valid for small-scale roughness of systems with evolving shear-bands. We note that similar to dislocations~\cite{zaiser_self-affine_2004,hinkle_emergence_2020}, a network of shear bands forms a network of steps on a surface, which also carries the signature of self-affine scaling with an exponent that depends on the lateral correlation of these steps. This indicates that during the formation of real-world surfaces, a number of mechanisms may be active at different scales that all lead to self-affine geometries.

\emph{Acknowledgements.} We thank Richard Leute, Gianpietro Moras, Laurent Ponson and Michael Zaiser for useful discussion. We used \textsc{lammps}~\cite{plimpton_fast_1995} for all calculations and \textsc{ovito pro}~\cite{ovito} and \textsc{pyvista} \cite{pyvista} for postprocessing and visualization. Topography data was analyzed with \textsc{contact.engineering}. We acknowledge support from the European Research Council (StG-757343) and the Deutsche Forschungsgemeinschaft (DFG, grant PA 2023/2). Calculations were carried out at the Jülich Supercomputing Center (JUWELS, grant hka18) and postprocessed at the University of Freiburg (NEMO, DFG grant INST 39/963-1 FUGG). Data is stored on bwSFS (University of Freiburg, DFG grant INST 39/1099-1 FUGG).


%

\clearpage

\setcounter{figure}{0}
\setcounter{equation}{0}

\renewcommand{\thesection}{S-\Roman{section}}
\renewcommand{\thefigure}{S-\arabic{figure}}
\renewcommand{\thetable}{S-\arabic{table}}
\renewcommand{\theequation}{S-\arabic{equation}}

\begin{center}
\Large\bf{ Supplemental Material for \\
 ``Nonequilibrium plastic roughening of metallic glasses yields self-affine topographies with strain-rate and temperature-dependent scaling exponents'' }
 \end{center}

\section{\hrms of a surface with a step formed by a shear band}
In Fig.~2(b), the curve for $T=\SI{10}{\kelvin}$ and $\dot\varepsilon=\SI{e7}{\per\second}$ 
does not scale $\propto\varepsilon^{0.5}$ at large strain. Shear bands are 
formed in this simulation, and the distinct $h_\mathrm{rms}$-curve can be described using
a simple model of the growth of a surface step generated by such a band.

Consider a line scan along the $x$-direction and assume that
the height profile created by the band can be described as
\begin{align}
h(x) = a \left(x - \frac{L(\varepsilon)}{2}\right),
\end{align}
where the dimensionless parameter $a$ describes the height of the step.
The mean height is zero, hence the $\hrms=L^{-1}\int_0^L
h^2(x)dx=aL/\sqrt{12}$. The length decreases with strain as
$L(\varepsilon)=\left(1-\varepsilon\right)L_0$. We assume that the
step height is a linear function of strain, i.e.\
\begin{align}
a(\varepsilon) = A\left(\varepsilon-\varepsilon_b\right) + B,
\end{align}
where $A$ and $B$ are constants, and $\varepsilon_y$ is the strain
at which the band is formed. In reality, the band forms over a
range of strain. However, this range is narrow, as can be seen in
Fig.~2(b), hence $\varepsilon_b$ is a good approximation.
In Fig.~2(b) $\varepsilon_b\approx{}\SI{4.8e-2}{}$.
At $\varepsilon_b$, \hrms jumps to a base value
$\hrms(\varepsilon_b)\equiv{}h_b\approx\SI{4}{\nano\meter}$,
which determines $B$.
By plugging $L(\varepsilon_b)$ and $a(\varepsilon_b)$ into
the formula for \hrms, we obtain
$B=h_b\sqrt{12}/\left(\left(1-\varepsilon_b\right)L_0\right)$, hence
\begin{align}
\hrms(\varepsilon) = \frac{L_0}{\sqrt{12}}
\left(A\left(\varepsilon-\varepsilon_b\right)+
\frac{h_b\sqrt{12}}{\left(1-\varepsilon_b\right)L_0}\right)\left(1-\varepsilon\right).
\end{align}
With  $A=1$, we obtain the red curve in Fig.~2(b), which
is close to the data for $\dot\varepsilon=\SI{1e7}{\per\second}$ and
$T=\SI{10}{\kelvin}$.

\end{document}